\documentclass[runningheads]{llncs}
\usepackage[T1]{fontenc}
\usepackage{graphicx}
\usepackage{cite}  
\usepackage{booktabs}
\usepackage{enumitem}
\usepackage{amsmath,amssymb,mathtools}
\usepackage{ulem}
\usepackage{listings}
\usepackage{hyperref}
\usepackage{algorithm}
\usepackage{listings}
\usepackage{algpseudocode}
\usepackage[most]{tcolorbox}
\usepackage{caption}
\usepackage{tikz}
\usetikzlibrary{arrows.meta,positioning}
\usepackage[htt]{hyphenat}
\usepackage[most]{tcolorbox}
\usepackage{enumitem}
\usepackage{ragged2e}
\usepackage{amsmath}
\usepackage[most]{tcolorbox}
\tcbset{
  macaCard/.style={
    enhanced, breakable,
    colback=gray!3, colframe=gray!55,
    colbacktitle=gray!12, coltitle=black, 
    boxrule=0.4pt, arc=1.5mm,
    left=1mm, right=1mm, top=0.6mm, bottom=0.6mm
  }
}

\usepackage{booktabs,siunitx,tabularx}
\newcolumntype{Y}{>{\raggedright\arraybackslash}X} 

\tcbset{
  listing engine=listings, listing only,
  colback=gray!3, colframe=black!30, boxrule=0.4pt, arc=2mm,
  left=1mm, right=1mm, top=1mm, bottom=1mm
}

\begin{document}
\title{
\textbf{MACA: A Framework for Distilling Trustworthy LLMs into Efficient Retrievers}
}
\titlerunning{MACA: Metadata-Aware Cross-Model Alignment}

\author{
  Satya Swaroop Gudipudi\inst{1} \and
  Sahil Girhepuje\inst{1} \and
  Ponnurangam Kumaraguru\inst{2} \and
  Kristine Ma\inst{1}
}

\authorrunning{S. Gudipudi et al.}

\institute{
  JP Morgan Chase \and
  IIIT Hyderabad
  \email{satyaswaroop.gudipudi@chase.com, sahil.girhepuje@chase.com, pk.guru@iiit.ac.in, kristine.ma@chase.com}
}



%
%
%
%
%
\maketitle              

\begin{abstract}
Modern enterprise retrieval systems must handle short, underspecified queries
such as ``foreign transaction fee refund'' and ``recent check status''. In these cases, semantic nuance and metadata matter but per-query large language model (LLM)
re-ranking and manual labeling are costly. We present Metadata-Aware Cross-Model Alignment (MACA), which distills a calibrated metadata aware LLM
re-ranker into a compact student retriever, avoiding online LLM calls. A
metadata-aware prompt verifies the teacher's trustworthiness by checking
consistency under permutations and robustness to paraphrases, then supplies
listwise scores, hard negatives, and calibrated relevance margins. The student trains with MACA's MetaFusion objective, which combines a metadata conditioned ranking loss with a cross model margin loss so it learns to push the correct answer above semantically similar candidates with mismatched topic, sub-topic,
or entity. On a proprietary consumer banking FAQ corpus and BankFAQs, the MACA teacher
surpasses a MAFA baseline at Accuracy@1 by five points on the proprietary set
and three points on BankFAQs. MACA students substantially outperform
pretrained encoders; e.g., on the proprietary corpus MiniLM Accuracy@1
improves from 0.23 to 0.48, while keeping inference free of LLM calls and
supporting retrieval-augmented generation.

\end{abstract}
\section{Introduction}

Modern enterprise retrieval systems must balance accuracy and latency while
handling underspecified user queries such as ``foreign transaction fee
refund'' or ``recent check status''. These queries are short and only partially specify their metadata; they hint at what the user wants but often omit key details about topic, sub-topic, intent, or entity. As a result, several candidate answers with different labels in the metadata taxonomy remain plausible. Conventional semantic retrievers may, for example, mix up foreign
transaction fees on credit cards with foreign ATM fees on debit cards, or
confuse queries like ``check the transaction status'' with ``recent check
status'', because they rely mainly on surface text similarity and make
limited use of structured metadata on the candidate side.

\begin{figure}[t]
  \centering
  \includegraphics[width=\linewidth]{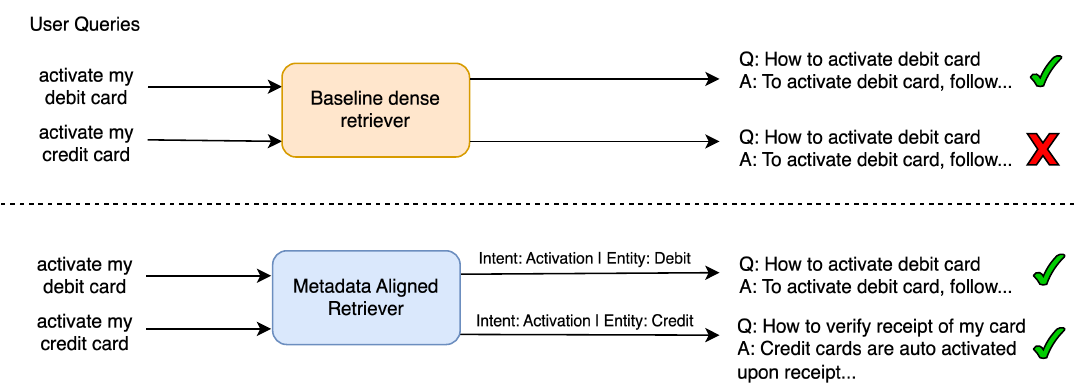}
  \caption{MACA reduces entity drift in short banking queries. The baseline
  returns the same debit-card FAQ for both queries; MACA selects the correct
  debit/credit FAQ at rank~1 (no LLM at inference).}
  \label{fig:maca_fig1}
\end{figure}

As illustrated in Fig.~\ref{fig:maca_fig1}, a baseline dense retriever may
route both ``activate my debit card'' and ``activate my credit card'' to the
same debit card FAQ. A metadata-aware retriever instead uses intent and card
type to select different answers for each query, without calling a large
language model (LLM) at inference time.

Production pipelines therefore rely on multi-stage retrieval: a lexical model
such as BM25~\cite{robertson2009bm25} first generates candidates, and a neural
re-ranker such as a cross-encoder, ColBERT, or an LLM re-ranker refines the
ordering to capture finer semantic distinctions~\cite{nogueira2019monot5,
khattab2020colbert,pradeep2023rankzephyr}. In enterprise settings, these
refinements are often needed to resolve subtle differences in user intent,
product, or fee type (e.g., distinguishing foreign transaction fees from
foreign ATM fees). However, per-query LLM inference is costly, latency-sensitive,
and operationally brittle~\cite{pradeep2023rankzephyr}. Large-scale human
labeling is also expensive and inconsistent, particularly when latent metadata
such as intent, topic, sub-topic, and entities must be inferred reliably.

Prior systems commonly use hybrid retrieval with LLM re-rankers~\cite{nogueira2019monot5,khattab2020colbert} or generic distillation of large
teachers into smaller students~\cite{hinton2015distillation,sanh2019distilbert}, but they tend to underuse explicit domain metadata. In
settings with large taxonomies, naively injecting all metadata into an LLM
prompt can exceed context limits and dilute relevance reasoning. Effective
retrieval therefore demands supervision that highlights the most salient
metadata signals without overwhelming the model.

We address these gaps with \textbf{MACA} (Metadata-Aware Cross Model
Alignment), a two-phase framework that distills a calibrated metadata-aware
LLM re-ranker into a compact, single pass retriever. In Phase~I, a
metadata-aware prompt first verifies the teacher's trustworthiness by
checking consistency under candidate permutations and robustness to
paraphrases. It then yields listwise scores, calibrated relevance margins, and
hard negatives where competing FAQs share wording but differ in their
structured metadata. In Phase~II, a compact MACA student learns from these
signals via a MetaFusion objective that combines (i) a metadata conditioned
ranking loss and (ii) a cross model margin loss (RCMA) that transfers the
teacher's margins over near-miss competitors. At inference time, only the
student runs, achieving low latency and stable quality with no online LLM
calls.

On a proprietary consumer banking FAQ dataset and the public BankFAQs dataset~\cite{public_bankfaqs_dataset}, the
MACA teacher surpasses a MAFA~\cite{hegazy2025mafa} agentic baseline at Accuracy@1 by about five
points on the proprietary set and three points on BankFAQs. MACA-distilled
students then substantially outperform pretrained encoders; for example, on
the proprietary corpus MiniLM Accuracy@1 improves from 0.23 to 0.48, while
remaining close to the teacher. This keeps inference free of LLM calls, which aligns well with the latency and governance constraints of retrieval-augmented generation in regulated domains.

\paragraph{Contributions.}
Our contributions are threefold:
(i) a trustworthiness protocol and metadata-aware prompt that calibrate an
LLM re-ranker along the axes of performance, consistency, and paraphrase-robustness;
(ii) the MACA MetaFusion objective, which combines metadata-conditioned
ranking with cross-model margin alignment on near-miss negatives selected by
a MACA Judge; and (iii) an empirical study on two banking FAQ corpora
showing that a calibrated LLM teacher can supervise compact retrievers that
recover most of the teacher’s gains without any per-query LLM cost.

\section{Related Work}

\paragraph{Multi-stage retrieval and fusion.}
Modern IR stacks typically combine a high-recall first-stage retriever with a
stronger re-ranker~\cite{nogueira2019bert}. Lexical BM25~\cite{robertson2009bm25}
remains a robust sparse baseline, while late interaction models such as
ColBERT~\cite{khattab2020colbert} improve effectiveness at manageable latency.
Benchmarks such as BEIR~\cite{thakur2021beir} document the trade-off between
effectiveness and latency across sparse, dense, and re-ranking families, while
Reciprocal Rank Fusion (RRF)~\cite{cormack2009rrf} often yields further gains
by aggregating heterogeneous candidates. Multi-stage BERT-based
rankers~\cite{nogueira2019bert} extend this pattern by stacking dense and
cross-encoder components. MACA follows this hybrid design in Phase~I, but
ultimately replaces the re-ranking stage with a single distilled retriever.

\paragraph{LLMs as re-rankers and judges.}
Prompted LLMs have emerged as strong zero and few-shot re-rankers: pairwise
ranking prompting (PRP)~\cite{qin2023prp}, RankGPT~\cite{sun2023rankgpt}, and
RankZephyr~\cite{pradeep2023rankzephyr} show that pairwise and listwise prompting
can match or surpass supervised cross-encoders on BEIR-style benchmarks. At
the same time, the LLM-as-a-judge literature, for example MT-Bench and related
work~\cite{mtbench}, and studies such as Can We Trust LLMs as Relevance
Judges?~\cite{trust_llm_judges} report that LLM judgments can be sensitive to
prompt wording, candidate order, and domain shift. This is especially true when they are used as a replacement for human assessors. More recent work on robustness of
benchmark-based LLM evaluation under paraphrasing finds that absolute scores
can degrade markedly even when model rankings remain relatively
stable~\cite{llm_robustness}. MACA is aligned with this line of work: instead
of assuming LLM re-rankers are reliable by default, we explicitly audit
candidate prompts using Accuracy@k, paraphrase Robustness@k, and order
Consistency@k, and we freeze only the most stable metadata-aware prompt as the
teacher.

\paragraph{Guided distillation for efficient retrievers.}
Distillation compresses strong but slow teachers into lightweight
students~\cite{hinton2015distillation}. For neural ranking, MarginMSE aligns
student and teacher margins and consistently improves efficient
rankers~\cite{hofstatter2020marginmse}; RocketQA~\cite{qu2021rocketqa} integrates hard-negative
mining and denoising for dense retrieval. Beyond
ranking, LLM-generated supervision has propelled instruction tuning and data
augmentation: Self-Instruct~\cite{wang2022selfinstruct}, Alpaca~\cite{taori2023alpaca},
and AugGPT~\cite{dai2023auggpt}. Our work builds on this line by (i)
calibrating a metadata-aware LLM prompt for reliability prior to labeling and
(ii) aligning a compact student to the prompt’s preferences via cross-model
margin objectives.

\paragraph{Metadata-aware guidance and multi-agent annotation.}
Metadata can boost accuracy but often increases latency. MOGIC, a recent
method for extreme classification, shows how an early-fusion metadata
``oracle'' can guide low-latency students with no serving-time
cost~\cite{prabhu2025mogic}. Closer to our setting, MAFA~\cite{hegazy2025mafa} uses multiple
specialized agents plus a judge for FAQ annotation and reports strong gains
on proprietary and public banking datasets. MACA differs by
eliminating per-query LLM calls at inference time: we first evaluate and
select a metadata-aware prompt for labeling, then distill its judgments into
a single student retriever aligned via cross-model margin learning.

\section{Methodology (MACA)}
\begin{figure}[t]
  \centering
  \includegraphics[width=\linewidth]{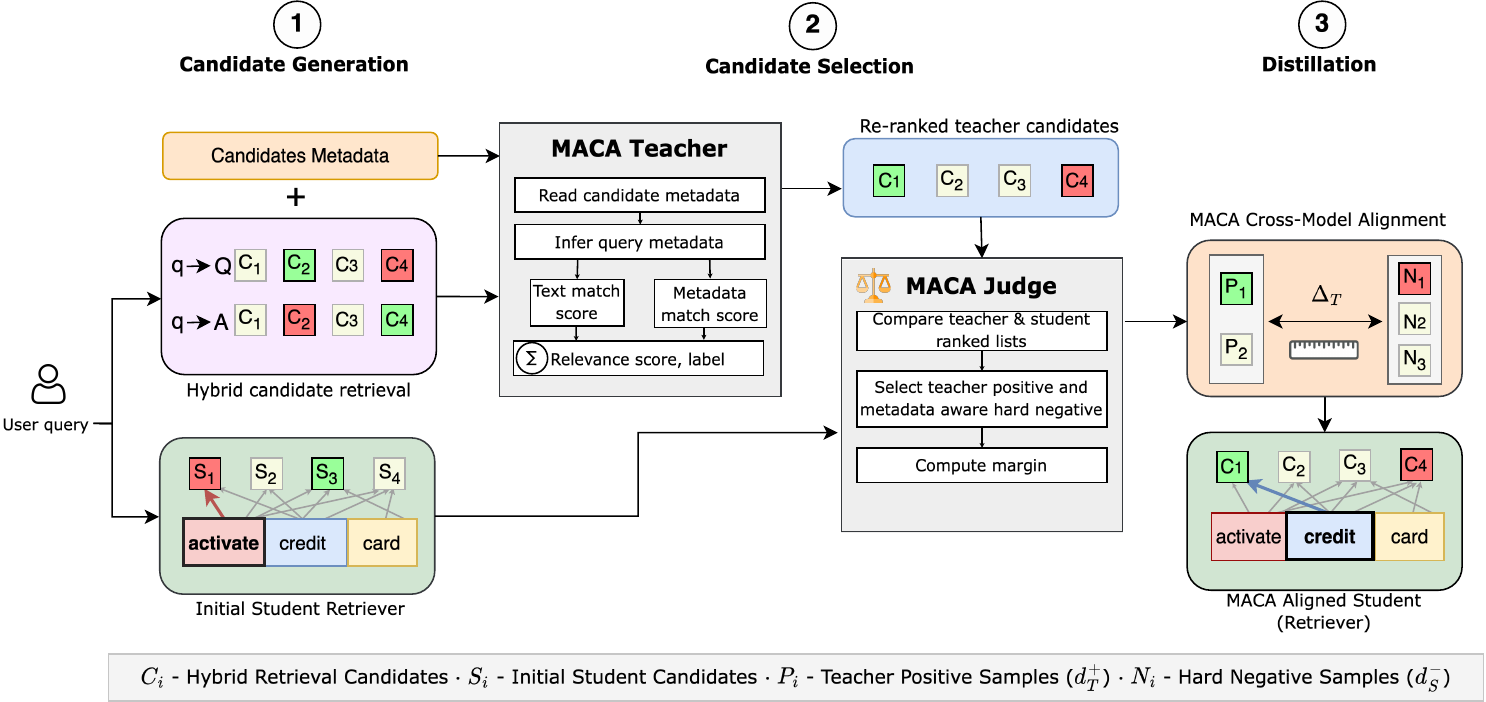}
  \caption{\textbf{MACA pipeline.} Phase I: candidates from the query-to-question view $q{\to}Q$ and the
query-to-answer view $q{\to}A$ are fused by RRF and re-ranked by a calibrated
MACA Teacher using metadata from the automatic taxonomy creation step
(Algorithm~\ref{alg:taxonomy}). Phase II: the MACA Judge selects a teacher
positive $d_T^{+}$ and a student hard negative $d_S^{-}$, and emits a margin
$\Delta_T$. These triplets are then used to train a compact student with
MNRL and RCMA. At inference time, only the student is used.}
  \label{fig:maca_fig2}
\end{figure}

We address two questions: (RQ1) can an LLM be calibrated for reliable,
metadata-aware relevance scoring, and (RQ2) can such a teacher supervise a
compact retriever so that inference requires no online LLM calls? Each item $f \in \mathcal{F}$ consists of a question, answer, and taxonomy
labels (topic, sub-topic, intent, entities) induced by
Algorithm~\ref{alg:taxonomy}. User queries are typically short and
ambiguous, making fine-grained human relevance labels expensive.
Figure~\ref{fig:maca_fig2} provides an overview of the MACA pipeline and the
interaction between Phase~I and Phase~II.

\subsection{Automatic taxonomy creation and labeling}
We use an LLM to induce a metadata taxonomy over topic, sub-topic, intent,
and entities, and to label queries and FAQ passages. Given a consumer banking
prior, the LLM first assigns a coarse topic to each unlabeled query. A second
pass, conditioned on this topic, predicts a sub-topic, intent, and salient
entities. The resulting
$(\text{topic},\text{sub-topic},\text{intent},\text{entity})$ tuples form an
automatic taxonomy, which we reuse to label FAQ questions and answers with
the same prompting scheme. In our experiments, we rely entirely on this
automatic taxonomy. In a production system, it could optionally be refined
once by subject-matter experts (SMEs). Algorithm~\ref{alg:taxonomy} summarizes
the procedure.

\begin{algorithm}[t]
\caption{Automatic LLM-based taxonomy creation and labeling}
\label{alg:taxonomy}
\begin{algorithmic}[1]
\Require Query set $Q$, FAQ set $\mathcal{F}$, domain prior $D$ (consumer banking), LLM $L$
\Ensure Taxonomy $\mathcal{T}$ over (topic, sub-topic, intent, entities) and metadata labels for $Q$ and $\mathcal{F}$

\State $\mathcal{T} \gets \emptyset$

\Comment{Step 1: assign topics to queries}
\ForAll{$q \in Q$}
  \State $t_q \gets L(D, q)$
  \State Store $t_q$ for $q$; update $\mathcal{T}.\mathrm{topics}$
\EndFor

\Comment{Step 2: fine-grained metadata for queries}
\ForAll{$q \in Q$}
  \State $(s_q, i_q, E_q) \gets L(D, q, t_q)$
  \State Store $(t_q, s_q, i_q, E_q)$ for $q$
  \State Update $\mathcal{T}$ with $(t_q, s_q, i_q, E_q)$
\EndFor

\Comment{Step 3: label FAQs using induced taxonomy}
\ForAll{$(Q_f, A_f) \in \mathcal{F}$}
  \State $(t_f, s_f, i_f, E_f) \gets L(\mathcal{T}, Q_f, A_f)$
  \State Store $(t_f, s_f, i_f, E_f)$ for FAQ $f$
  \State Update support counts for $(t_f, s_f, i_f, E_f)$ in $\mathcal{T}$
\EndFor

\Comment{Step 4 (optional): SME refinement for production systems}
\State Optionally refine $\mathcal{T}$ by merging near-duplicate labels and pruning low-support entries
\State \Return $\mathcal{T}^*$ and all metadata labels for $Q$ and $\mathcal{F}$
\end{algorithmic}
\end{algorithm}

\subsection{Phase I: Calibrated LLM re-ranker (RQ1)}
Phase~I builds a metadata aware LLM re-ranker that serves as the teacher.
For MACA to be useful, this teacher must be trustworthy: its
judgments should be accurate on held-out labels (performance), stable under
benign changes to the candidate list (consistency), and insensitive to
paraphrases of the same query (robustness). A teacher that returns different
top answers for ``card activation'' and ``activate card'' would simply bake
instability into the student, so we first calibrate and select a prompt that
scores well on all three dimensions, and only then freeze this teacher for
distillation.

\paragraph{Candidate generation.}
For each query $q$, we build a high-recall candidate set from two dense
views over the FAQ corpus. The query-to-question view $q{\to}Q$ searches
only FAQ questions, and the query-to-answer view $q{\to}A$ searches only FAQ
answers. $q{\to}Q$ works best when queries resemble FAQ titles, while
$q{\to}A$ improves coverage when crucial details (e.g., card type or foreign
transaction fees) are spelled out in the answer text rather than the
question. We rank each view separately and fuse them with Reciprocal Rank
Fusion (RRF)~\cite{cormack2009rrf}; the fused top-$K$ candidates are then
passed to the LLM re-ranker.

\paragraph{Metadata-aware re-ranking.}
A metadata-aware prompt asks the LLM to (i) infer query-level metadata
(intent, topic, sub-topic, entities) and (ii) score each candidate using both
text and metadata alignment. The LLM assigns TREC-style four-point grades
(exact, partial, less relevant, irrelevant), which we map to numerical labels
$3,2,1,0$ following TREC DL~\cite{craswell2021trecdl}. In addition, it
produces a continuous relevance score $S_T(d)$ for each candidate $d$, based
on a weighted combination of text coverage and metadata overlap (intent,
topic, sub-topic, entity Jaccard); these scores later define margins between
positives and near-miss competitors.

We explore several prompting strategies, including chain-of-thought,
Reflexion, and ReAct~\cite{wei2022chainofthought,%
reflexion2023,yao2022react}, and measure their
trustworthiness as a three-dimensional profile over Accuracy@k,
Consistency@k, and Robustness@k (Section~\ref{sec:trustworthiness}). We then
select, for RQ1, the prompt with the best trade-off across these three axes
and freeze it as the MACA teacher.

\subsection{Phase II: Candidate selection and distillation (RQ2)}

\paragraph{Candidate selection (MACA Judge).}
Given a query $q$, the frozen teacher re ranker returns a top $k_T$ list with scores $S_T(\cdot)$ and the student retriever returns a top $k_S$ list with scores $S_S(\cdot)$. The MACA Judge is an LLM prompt that sees both lists with metadata and teacher labels, and selects a positive and hard negative pair.

(1) Teacher positive. The judge sets $d_T^{+}$ to the highest scoring teacher candidate labeled exact, or if none exists, the highest scoring partial.

(2) Metadata-aware hard negative. The judge then forms a set of near miss student candidates
\begin{equation}
\begin{aligned}
\mathcal{N} = \{\, d \in \text{student top }k_S :\ 
& d \neq d_T^{+},\ d \text{ not marked relevant by the judge},\\
& \mathrm{overlap}(d_T^{+}, d) \ge 1 \,\},
\end{aligned}
\label{eq:near-miss-set}
\end{equation}
where $\mathrm{overlap}$ tests for shared topic or intent and allows sub-topic or entity differences (for example foreign transaction fee versus foreign ATM fee).
If $\mathcal{N} = \varnothing$, we fall back to the highest scoring student item that the judge marks non relevant. The hard negative is
\begin{equation}
d_S^{-} = \arg\max_{d\in\mathcal{N}} S_S(d)
\label{eq:hard-negative}
\end{equation}

(3) Calibrated margin. For $(d_T^{+}, d_S^{-})$, the judge uses teacher scores to compute
\begin{equation}
\Delta_T = S_T(d_T^{+}) - \max\!\big(S_T(d_S^{-}),\, \gamma\big),
\label{eq:teacher-margin}
\end{equation}
with a label aware floor $\gamma$ and clipping to $[-m_{\max}, m_{\max}]$ to curb outliers.

\paragraph{Distillation.}
The student is trained with
\begin{equation}
\mathcal{L}_{\text{total}} = \alpha\,\mathcal{L}_{\text{MNRL}} + \beta\,\mathcal{L}_{\text{RCMA}}
\label{eq:total-loss}
\end{equation}
Here, $\mathcal{L}_{\text{MNRL}}$ is the standard in batch Multiple Negatives Ranking Loss~\cite{henderson2017smartreply}, and RCMA is a MarginMSE style distillation term~\cite{hofstatter2020marginmse}:
\begin{equation}
\Delta_S = S_S(q, d_T^{+}) - S_S(q, d_S^{-}), \qquad 
\mathcal{L}_{\text{RCMA}} = \big(\Delta_S - \Delta_T\big)^2
\label{eq:rcma}
\end{equation}
This alignment transfers the teacher separation between positive and hard negative, encouraging the student to reproduce metadata aware gaps.

In summary, MACA consists of an automatic LLM-based metadata induction step
(Algorithm~\ref{alg:taxonomy}), a calibrated metadata-aware LLM re-ranker that
scores candidates (Phase~I), and a MACA Judge that selects positives, hard
negatives, and margins to train a compact student via MetaFusion (Phase~II).
At inference time, only the student retriever is used.

\section{Experiments}

\subsection{Setup}
\paragraph{Datasets}
We use two FAQ corpora. (i) Proprietary consumer banking: 15k unlabeled
queries for distillation and 250 expert-labeled queries for evaluation; its
metadata taxonomy ($\sim$120 sub-topics and $\sim$150 entities) is induced
automatically using Algorithm~\ref{alg:taxonomy} and used as enums in prompts
and for metadata-aware evaluation. (ii) BankFAQs: 2{,}400 question–answer
pairs; we paraphrase questions with \texttt{gpt-4o} to obtain 2{,}400 synthetic
training queries and a 250-query held-out set.

\paragraph{Teacher and students}
Phase~I (teacher) uses dense $q{\to}Q$ and $q{\to}A$ retrieval with OpenAI
\texttt{text-embedding-3-large}, fused with RRF~\cite{cormack2009rrf}, then a
metadata-aware \texttt{gpt-4o} re-ranker (the MACA teacher, used only offline
for labeling). Phase~II distills into three off-the-shelf encoders widely
used in production retrieval:
\texttt{all-MiniLM-L6-v2}, \texttt{all-mpnet-base-v2}, and
\texttt{msmarco-distilbert-base-tas-b}~\cite{wang2020minilm,mpnet2020,distiltasb2021}.
We refer to these as MiniLM, MPNet, and TAS-B in
Table~\ref{tab:main}. We report models fine-tuned with MACA–MNRL and
MACA–MetaFusion (MNRL+RCMA).

\paragraph{Metrics and trustworthiness}\label{sec:trustworthiness}
\textbf{Accuracy@k} is the fraction of queries for which at least one relevant
candidate appears in the top-$k$ results ($k \in \{1,3,5,10,15\}$).

For each model we report a three-dimensional trustworthiness profile over
Accuracy@k, Consistency@k (C@k), and Robustness@k (R@k). Let $Q$ be the set of
evaluation queries. For permutations, let $\mathrm{topk}_p(q)$ be the top-$k$
set for query $q$ under permutation $p$ of a fixed candidate list, with $P_q$
permutations per query. For paraphrases $\{q^{(i)}\}_{i=1}^{M_q}$ of $q$, let
$\mathrm{topk}(q^{(i)})$ be the top-$k$ set for paraphrase $i$. We define
\begin{equation}
\begin{aligned}
C@k &= \frac{1}{|Q|}\sum_{q\in Q} \frac{1}{P_q}
       \max_{S}\#\{\,p : \mathrm{topk}_p(q)=S\,\}, \\
R@k &= \frac{1}{|Q|}\sum_{q\in Q} \frac{1}{M_q}
       \max_{S}\#\{\,i : \mathrm{topk}(q^{(i)})=S\,\}.
\end{aligned}
\label{eq:trustworthiness}
\end{equation}
Thus C@k is the average fraction of permutations that agree with the modal
top-$k$ set, and R@k is the corresponding fraction over paraphrases.

\paragraph{Training}
Unless noted, students train for 4 epochs with Adam (learning rate
$2\times10^{-5}$, batch size 64, linear warmup); MetaFusion weights
$(\alpha,\beta)$ are selected on the validation set. Code release is
subject to organizational approval.

\subsection{Results}
\begin{figure}[t]
  \centering
  \includegraphics[width=0.95\columnwidth]{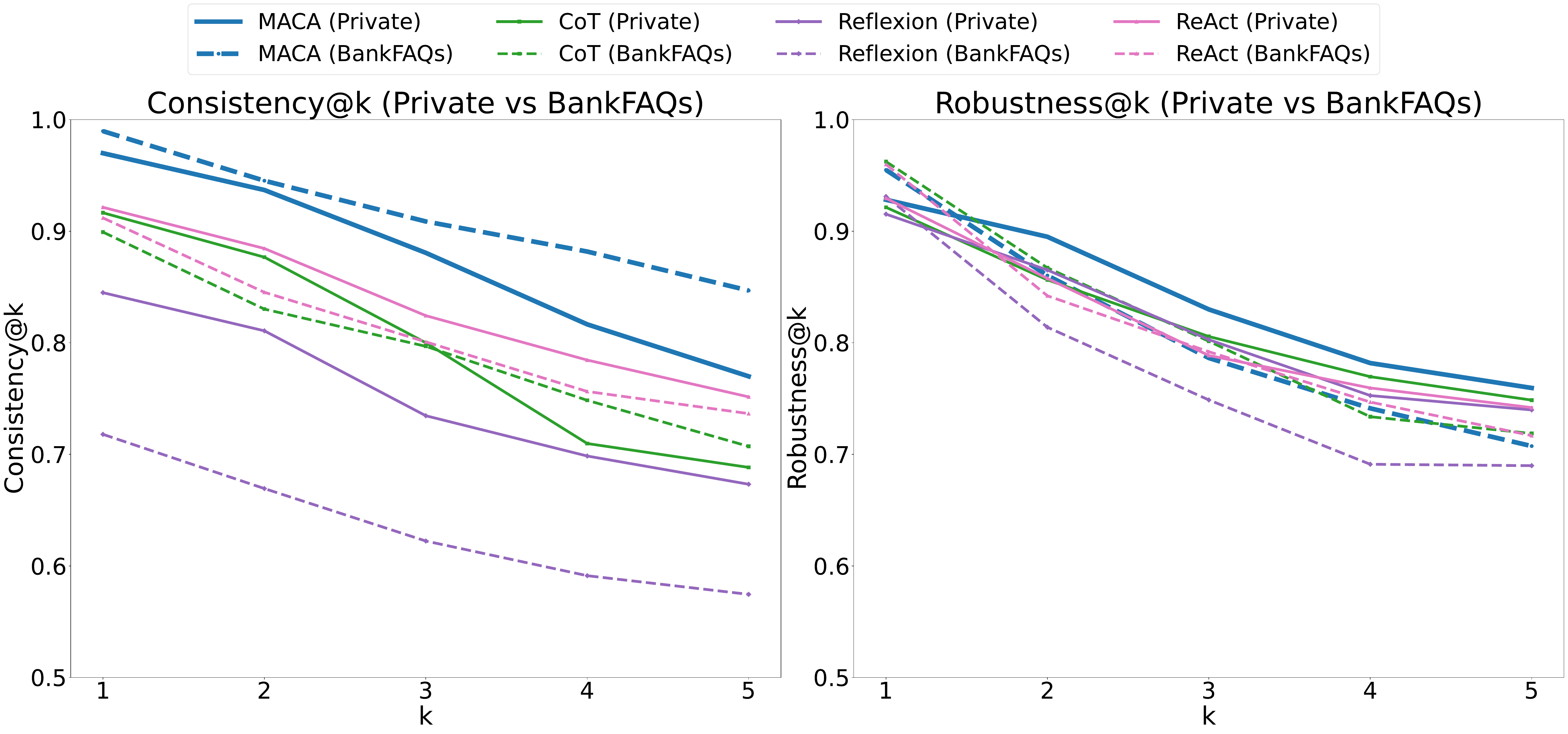}
  \caption{Trustworthiness (Consistency@1, Robustness@1) of four prompts on both datasets; MACA is the most balanced.}
  \label{fig:radar-two}
\end{figure}

\paragraph{RQ1: Trustworthy LLM teachers.}
Table~\ref{tab:prompts} compares LLM prompting strategies when used as
re-rankers. On the proprietary banking set, MACA attains the best accuracy
at all cutoffs (Acc@1/3/5 = 0.61/0.79/0.84), improving over CoT and
Reflexion by 3--4 points at Acc@1 and over MAFA by 5 points. On BankFAQs,
which is easier because queries are paraphrases of FAQ questions, all
prompts reach high accuracy, but MACA still gives the strongest top-1
performance (Acc@1 = 0.72 vs.\ 0.71 for CoT/ReAct and 0.66--0.68 for
Reflexion/MAFA) while matching the best Acc@3/5. Fig.~\ref{fig:radar-two}
shows a similar pattern for stability: across $k\!\in\!\{1,\dots,5\}$ MACA
has the highest Consistency@k and remains the most robust to paraphrasing,
with smaller gaps between private and BankFAQs curves than CoT, ReAct, or
Reflexion. Overall, MACA offers the best trade-off between performance,
consistency, and robustness, supporting its use as the frozen teacher for
distillation.
\begin{table}[t]
\footnotesize
\setlength{\tabcolsep}{1.5pt}
\centering
\caption{Prompt-level ablation on Consumer Banking (proprietary) and BankFAQs:
Accuracy@\{1,3,5\} for different LLM prompting strategies.}
\label{tab:prompts}
\begin{tabular*}{\columnwidth}{@{\extracolsep{\fill}}l
  *{3}{S[table-format=1.2]} *{3}{S[table-format=1.2]}@{}}
\toprule
& \multicolumn{3}{c}{Consumer Banking (proprietary)}
& \multicolumn{3}{c}{BankFAQs} \\
\cmidrule(lr){2-4}\cmidrule(lr){5-7}
Prompt & {Acc@1} & {Acc@3} & {Acc@5}
       & {Acc@1} & {Acc@3} & {Acc@5} \\
\midrule
CoT (chain-of-thought)  & 0.57 & 0.73 & 0.79 & 0.71 & 0.88 & 0.92 \\
ReAct                   & 0.53 & 0.74 & 0.77 & 0.71 & 0.89 & 0.92 \\
Reflexion               & 0.58 & 0.73 & 0.81 & 0.66 & 0.87 & 0.90 \\
MACA (ours)             & 0.61 & 0.79 & 0.84 & 0.72 & 0.89 & 0.92 \\
MAFA                    & 0.56 & 0.70 & 0.77 & 0.68 & 0.81 & 0.86 \\

\bottomrule
\end{tabular*}
\end{table}

\begin{table}[t]
\footnotesize
\setlength{\tabcolsep}{1.5pt}
\centering
\caption{Accuracy@\{1,5,10,15\} for baselines and MACA-distilled students on
Consumer Banking (proprietary) and BankFAQs. The MACA teacher (LLM re-ranker) is used
only offline to label training data; inference uses the student retrievers.
MiniLM, MPNet, and TAS-B denote \texttt{all-MiniLM-L6-v2},
\texttt{all-mpnet-base-v2}, and \texttt{msmarco-distilbert-base-tas-b},
respectively; “+MNRL” and “+MF” indicate MACA fine-tuning with MNRL and
MACA–MetaFusion.}

\label{tab:main}
\begin{tabular}{@{}l *{4}{S[table-format=1.2]} *{4}{S[table-format=1.2]}@{}}
\toprule
& \multicolumn{4}{c}{Consumer Banking (proprietary)}
& \multicolumn{4}{c}{BankFAQs} \\
\cmidrule(lr){2-5}\cmidrule(lr){6-9}
Model & {Acc@1} & {Acc@5} & {Acc@10} & {Acc@15}
      & {Acc@1} & {Acc@5} & {Acc@10} & {Acc@15} \\
\midrule
BM25                    & 0.19 & 0.44 & 0.51 & 0.55 & 0.38 & 0.65 & 0.71 & 0.75 \\
MiniLM                  & 0.23 & 0.61 & 0.71 & 0.79 & 0.62 & 0.90 & 0.93 & 0.95 \\
MiniLM+MNRL             & 0.46 & 0.77 & 0.80 & 0.83 & 0.69 & 0.93 & 0.96 & 0.97 \\
MiniLM+MF               & 0.48 & 0.77 & 0.82 & 0.83 & 0.70 & 0.94 & 0.96 & 0.97 \\
MPNet                   & 0.32 & 0.57 & 0.70 & 0.75 & 0.64 & 0.92 & 0.95 & 0.97 \\
MPNet+MNRL              & 0.44 & 0.76 & 0.78 & 0.81 & 0.70 & 0.94 & 0.96 & 0.97 \\
MPNet+MF                & 0.47 & 0.76 & 0.78 & 0.79 & 0.70 & 0.94 & 0.96 & 0.97 \\
TAS-B                   & 0.26 & 0.46 & 0.58 & 0.65 & 0.58 & 0.89 & 0.95 & 0.97 \\
TAS-B+MNRL              & 0.42 & 0.76 & 0.79 & 0.81 & 0.64 & 0.92 & 0.97 & 0.97 \\
TAS-B+MF                & 0.46 & 0.75 & 0.79 & 0.81 & 0.65 & 0.91 & 0.97 & 0.97 \\
\bottomrule
\end{tabular}
\end{table}

\paragraph{RQ2: Distilled retrieval quality.}
Table~\ref{tab:main} reports retrieval accuracy for three students distilled
from the MACA teacher. On the proprietary corpus, MACA fine-tuning
substantially improves all backbones over BM25 and their off-the-shelf
versions: Acc@1 rises from 0.23/0.32/0.26 for MiniLM/MPNet/TAS-B to
0.48/0.47/0.46 with MetaFusion, closing much of the gap to the teacher
(0.61). Similar gains appear at higher cutoffs, e.g., MiniLM Acc@5 improves
from 0.61 to 0.77, while MPNet and TAS-B show analogous trends. On
BankFAQs, where pretrained encoders already achieve Acc@1 $\geq$ 0.58 and
Acc@5 $\geq$ 0.89, MACA yields smaller but consistent lifts (e.g., MiniLM
Acc@1 from 0.62 to 0.70 and MPNet from 0.64 to 0.70). In both datasets,
MACA students recover a large fraction of the LLM teacher’s effectiveness
while requiring only a single encoder pass and no online LLM calls.

\paragraph{Ablation: MetaFusion loss.}
From Table~\ref{tab:main}, most of the gain over BM25 and the pretrained
students comes from MNRL, which roughly doubles Acc@1 on the proprietary set
(e.g., MiniLM 0.23$\to$0.46). MetaFusion (+MF) then adds a smaller but
consistent boost at the very top rank across all backbones, gaining 2--4
points at Acc@1 on Consumer Banking and about 1 point on BankFAQs, while
leaving Acc@5 nearly unchanged. This is consistent with RCMA’s design: by
aligning the student to the teacher’s margin between the positive and a
metadata-aware hard negative, MetaFusion focuses learning on borderline pairs
that determine strict top-$k$ retrieval. The MACA Judge’s near-miss selection
ensures that these pairs differ mainly in sub-topic or entity, so the student
learns to preserve metadata-sensitive distinctions that matter most for
top-ranked answers.

\section{Discussion and Future Work}
MACA was designed to test whether an LLM can be calibrated into a trustworthy
metadata-aware relevance teacher (RQ1) and whether such a teacher can
supervise compact retrievers that remove online LLM calls (RQ2). Our results
in the banking FAQ setting indicate that both goals are achievable: the MACA
prompt outperforms CoT/ReAct/Reflexion and the MAFA agentic baseline on
accuracy, consistency, and robustness, and MACA-distilled students
substantially outperform BM25 and off-the-shelf dense encoders while
recovering much of the teacher’s effectiveness.

These findings suggest practical benefits for retrieval and RAG in regulated
domains. A calibrated teacher plus MetaFusion yields single-pass students
that are fast yet metadata-aware, avoiding per-query LLM calls. In addition,
our trustworthiness metrics (Accuracy@k, Consistency@k, Robustness@k)
provide a concrete protocol for auditing LLM re-rankers before using them
for large-scale labeling, which is valuable for practitioners who must
justify model choices.

As future work, we plan to extend MACA beyond banking to domains such as
medical and legal search, and to incorporate richer metadata, for example
user or channel context. We also aim to broaden our notion of
trustworthiness to include calibration and uncertainty-aware labeling, and
to evaluate MACA in full RAG pipelines to measure its impact on
end-to-end answer quality.

\section*{Disclaimer}
The views and opinions expressed in this work are those of the authors
and do not necessarily reflect the official policies or positions of
their current or former employers.

\appendix
\section{MACA Prompts}

\begin{figure*}[htbp]
\centering
\begin{minipage}{0.49\textwidth}
\begin{tcolorbox}[macaCard, title={MACA Re-ranker Prompt (summary)}]
\RaggedRight\scriptsize

\textbf{Task.} Rank candidate FAQs for a query using provided taxonomy metadata only.

\textbf{Inputs.} \texttt{query}; \texttt{candidates} with \texttt{intent, topic, sub\_topic, entities}.

\textbf{Steps.}
\begin{enumerate}[leftmargin=*, itemsep=1pt, topsep=1pt]
  \item Infer query metadata: intent, topic, sub\_topic (slug $ \le 3 $ words or ``unknown''), entities ($ \le 5 $ tokens).

  \item For each candidate: compute intent/topic/subtopic matches; entity overlap $|\cap|$ and Jaccard $|\cap|/|\cup|$.
  \item Scores: coverage $\in[0,1]$ (minimal read; 0 if unsure).  
  metadata $=0.40\,[\mathrm{intent}]+0.25\,[\mathrm{topic}]+0.20\,[\mathrm{subtopic}]+0.15\times\mathrm{Jaccard}$.  
  relevance $=0.65\times\mathrm{coverage}+0.35\times\mathrm{metadata}$.
  \item Labels: \texttt{exact} if intent and (topic or subtopic) and overlap$\ge$1;  
  \texttt{partial} if any match but not exact;  
  \texttt{less\_relevant} if topic matches but main entity differs;  
  \texttt{irrelevant} otherwise.
  \item Ambiguity: if intent unclear, prefer actionable (how\_to/troubleshooting/policy\_limit/eligibility) when scores are close.
\end{enumerate}

\textbf{Output (strict JSON, top-1 shown).}
\begin{verbatim}
{"faq_id":"F123","label":"exact",
 "relevance_score":0.87,
 "top_matching_entity":"activation"}
\end{verbatim}
\end{tcolorbox}
\end{minipage}\hfill
\begin{minipage}{0.49\textwidth}
\begin{tcolorbox}[macaCard, title={MACA-Judge Prompt (summary)}]
\RaggedRight\scriptsize

\textbf{Task.} Select teacher positive + student hard negative; output calibrated margin.

\textbf{Inputs.} \texttt{query}; teacher top-$k_T$ (scores $S_T$); student top-$k_S$ (scores $S_S$); metadata.

\textbf{Policy.}
\begin{enumerate}[leftmargin=*, itemsep=1pt, topsep=1pt]
  \item \textbf{Positive} $d_T^{+}$: highest teacher item labeled exact; else best partial.
  \item \textbf{Near-miss set} $\mathcal{N}$: student items not marked relevant by teacher, sharing topic or intent with $d_T^{+}$; remove duplicates.
  \item \textbf{Hard negative} $d_S^{-}$: $\arg\max_{d\in\mathcal{N}} S_S(d)$; tie-break by metadata proximity (topic $>$ intent $>$ entity), then by $S_T(d)$.
  \item \textbf{Margin} $\Delta_T = S_T(d_T^{+})-\max(S_T(d_S^{-}),\gamma)$, clipped to $[-m_{\max},m_{\max}]$.
\end{enumerate}

\textbf{Output (strict JSON).}
\begin{verbatim}
{"query":"...","pos_id":"F123",
 "neg_id":"F987","delta_T":0.42}
\end{verbatim}
\end{tcolorbox}
\end{minipage}
\caption{Compact prompt cards for MACA. Left: metadata-aware re-ranker. Right: deterministic judge for triplet selection and margin calibration.}
\label{fig:prompt_cards_compact}
\end{figure*}

%
%
%
%

\end{document}